\documentclass[%
 reprint,twocolumn,
 amsmath,amssymb,
 aps,
]{revtex4-1}

\usepackage{graphicx}
\usepackage{dcolumn}
\usepackage{bm}
\usepackage{color}

\begin{document}

\title{Topology and the optical Dirac equation}

\author{S. A. R. Horsley}
\affiliation{Department of Physics and Astronomy,
University of Exeter, Stocker Road, Exeter EX4 4QL}

\date{\today}

\begin{abstract}
Through understanding Maxwell's equations as an effective Dirac equation (the `optical Dirac equation'), we re--examine the relationship between electromagnetic interface states and topology.  We illustrate a simple case where electromagnetic material parameters play the roles of `mass' and `energy' in an equivalent Dirac equation.  The modes trapped between a gyrotropic medium and a mirror are then the counterpart of those at a `domain wall', where the mass of the Dirac particle changes sign.  Considering the general case of arbitrary electromagnetic media, we provide an analytical proof relating the integral of the Berry curvature (the Chern number) to the number of interface states.  We show that this reduces to the usual result for periodic media, and also that the Chern number can be computed without knowledge of how the material parameters depend on frequency.
\end{abstract}

\maketitle
Dirac sought to resolve an inconsistency between relativity and quantum mechanics through looking for an operator whose \emph{square} gives the relativistic energy formula.  This led him to the following Hamiltonian~\cite{dirac1928} (in units where $\hbar=c=1$),
\begin{equation}
\hat{H}_{\rm D}=\boldsymbol{\alpha}\cdot\hat{\boldsymbol{p}}+m\beta\label{eq:Dirac_Hamiltonian}.
\end{equation}
where $\alpha_{i}$ and $\beta$ are matrices.  The Dirac equation is then given by $\hat{H}_{\rm D}|\psi\rangle={\rm i}\partial|\psi\rangle/\partial t$.  For example, in two spatial dimensions these matrices are given by,
\begin{equation}
\alpha_{x}=\left(\begin{matrix}0&1\\1&0\end{matrix}\right),\;\alpha_{y}=\left(\begin{matrix}0&-{\rm i}\\{\rm i}&0\end{matrix}\right),\;\beta=\left(\begin{matrix}1&0\\0&-1\end{matrix}\right)
\end{equation}
i.e. the three Pauli matrices.  For a fixed energy $\mathcal{E}$, the Dirac equation is then given by the following $2\times2$ matrix equation~\cite{shen2017}
\begin{equation}
	\left(\begin{matrix}\mathcal{E}-m&2{\rm i}\frac{\partial}{\partial z}\\2{\rm i}\frac{\partial}{\partial z^{\star}}&\mathcal{E}+m\end{matrix}\right)\left(\begin{matrix}\psi_{a}\\\psi_{b}\end{matrix}\right)=0\label{eq:dirac2d}
\end{equation}
where $z=x+{\rm i}y$, and the wave--function has two components given by $\psi_{a}$ and $\psi_{b}$.
\par
This equation (or rather its 4D cousin) has become a core part of quantum electrodynamics and is the foundation of the modern theory of the electron~\cite{weinberg1995,thaller2013}.  It can also be generalized to higher spin particles~\cite{rarita1941}, and an analogue holds for electromagnetic radiation~\cite{birula1994,barnett2014}.  Since its discovery, this equation has also appeared in some rather unexpected places, playing a role in spectral geometry~\cite{esposito1998}, and the associated Atiyah--Singer index theorem~\cite{atiyah1963,atiyah1973}, connecting the subject of topology to an analysis of the zero modes of a Dirac operator.  Special cases of these results appear in quantum mechanics~\cite{aharonov1979}, and have recently been investigated in non--Hermitian optics~\cite{leykam2017}.  An optical Dirac equation has also been applied to understand the division of electromagnetic spin and orbital angular momentum~\cite{barnett2014}.  A great deal is already known about the relationship between the Dirac equation and topology, and this paper represents an effort to translate these ideas into the optics of electromagnetic media.  Writing Maxwell's equations in the form of a Dirac equation will be shown to be very useful for understanding how topology can be applied to design optical materials.
\par
The main findings of this work can be summarized as follows.  We first give what we believe is the simplest example of a Dirac operator in optics: the two dimensional Dirac equation describes the propagation of electromagnetic waves in gyrotropic media, with the optical gyrotropy being analogous to the mass of the Dirac particle.  This shows the equivalence of unidirectional waves at the interface of such media to those that occur for the Dirac operator at a `domain wall' where the mass changes sign (which is the archetype of a topological insulator~\cite{hasan2010,shen2017}).  We then show Maxwell's equations can be more generally understood as a Dirac--like equation.  As an application of this analogy we prove a result, relating the number of these unidirectional interface states to a topological quantity (winding number) computed from the Green function.  While the use of the Dirac equation to describe electromagnetic interface states is commonplace in the physics of graphene analogues (see e.g.~\cite{bellec2013}), we believe that this is the first place that the relationship has been shown to hold for continuous media.
\par
Our findings are relevant to a large body of recent work, where the mathematics of topology utilized in condensed matter physics~\cite{niu1985,essin2011} has been carried over from electron waves to electromagnetic waves, giving rise to so--called `photonic topological insulators'~\cite{haldane2008,lu2014,bliokh2015}.  Our work is an effort to investigate the theory underlying the blossoming field of topological photonics.  Despite the early work by Haldane~\cite{haldane2008} and several recent papers by Silveirinha~\cite{silveirinha2015,silveirinha2016}, there has been relatively little done to examine the theoretical fundamentals in this field.  Here we find that topological quantities---such as the ubiquitous Chern number---can be computed in the space of material parameters at a \emph{fixed} frequency, something which has been noted before~\cite{wang2008,gao2015} without (as far as the author is aware) having been proven.  This twist on the existing method for designing photonic topological insulators is quite general, and can sometimes simplify calculations; avoiding having to invent a frequency dependence for the material, and thus also the introduction of dissipation~\cite{haldane2008,silveirinha2015}.This work also complements the recent findings of Van Mechelen and Jacob, who have similarly developed the parallel between the Maxwell and Dirac equations, although the application of their formalism is quite different~\cite{mechelen2017}.
\par
We begin with the simple case of electromagnetic waves in a homogeneous planar medium.  The general case of inhomogeneous media will be treated in the final section.  Consider a \emph{monochromatic} electromagnetic field confined to propagate in the $x$--$y$ plane, and polarized with $\boldsymbol{H}=H\hat{\boldsymbol{z}}$ (TM polarized).  When the plane is filled with an anisotropic medium of relative in--plane permittivity $\boldsymbol{\epsilon}$ and permeability $\mu=1$, Maxwell's equations reduce to
\begin{align}
\boldsymbol{\nabla}H\times\hat{\boldsymbol{z}}&=-{\rm i}\omega\epsilon_0\boldsymbol{\epsilon}\cdot\boldsymbol{E}\nonumber\\[10pt]
\boldsymbol{\nabla}\times\boldsymbol{E}&={\rm i}\omega\mu_0 H\hat{\boldsymbol{z}}.
\end{align}
We assume an in--plane permittivity tensor of Hermitian form (c.f.~\cite{davoyan2013,silveirinha2015,horsley2017})
\begin{align}
	\boldsymbol{\epsilon}&=\left(\begin{matrix}\epsilon_1&{\rm i}\alpha\\-{\rm i}\alpha&\epsilon_1\end{matrix}\right)\nonumber\\[10pt]
    &=\mathcal{E}\left(\mathcal{E}-m\right)\boldsymbol{e}_{+}\otimes\boldsymbol{e}_{-}+\mathcal{E}\left(\mathcal{E}+m\right)\boldsymbol{e}_{-}\otimes\boldsymbol{e}_{+}\label{eq:permittivity}
\end{align}
where the eigenvectors of this tensor are given by $\boldsymbol{e}_{\pm}=(\hat{\boldsymbol{x}}\pm{\rm i}\hat{\boldsymbol{y}})/\sqrt{2}$, and we have introduced the notation $\mathcal{E}=\sqrt{\epsilon}_{1}$ and $m=\alpha/\sqrt{\epsilon}_{1}$.  This kind of material is known as gyrotropic, and the quantity $\alpha$ is the magnitude of the gyration vector $\boldsymbol{g}=\alpha\hat{\boldsymbol{z}}$~\cite{volume8}.  In terms of the complex vector basis $\boldsymbol{e}_{\pm}$ the equations for the electric and magnetic fields can be written as
\begin{align}
	\frac{\partial\eta_0 H}{\partial z}&=-\frac{k_0\mathcal{E}(\mathcal{E}-m)}{\sqrt{2}}E_{+}\nonumber\\
    \frac{\partial\eta_0 H}{\partial z^{\star}}&=\frac{k_0\mathcal{E}(\mathcal{E}+m)}{\sqrt{2}}E_{-}\nonumber\\
    \frac{\partial E_{+}}{\partial z^{\star}}-\frac{\partial E_{-}}{\partial z}&=\frac{k_0\eta_0}{\sqrt{2}}H\label{eq:maxwell_complex}
\end{align}
where $\eta_0=\sqrt{\mu_0/\epsilon_0}$.  Written in terms of the complex coordinates $z,z^{\star}$, there is an interesting way to understand these equations.  The derivative of a function with respect to $z$ at a given point is a measure of how much the phase of $f$ winds anti--clockwise around that point.  In addition, $E_+$ is the part of the electric field vector that rotates in the plane in the anti--clockwise sense.  Equation (\ref{eq:maxwell_complex}) thus states that the anti--clockwise rotation of the phase of $H$ is proportional to the anti--clockwise rotation of the electric field vector, and equivalently for the clockwise rotation.  The non--zero value of $m$ biases the coupling between the two different directions of rotation of the phase of $H$ and those of the electric field.    In the extreme case where e.g. $\mathcal{E}=m$, the phase of the magnetic field is forced to rotate in only one direction.  This leads to the out of plane field $H$ becoming an analytic function of position, and thus exhibits unidirectional propagation.  This phenomenon occurs when the material parameters are on the tipping point between transparency and opacity, and can be associated with a zero of the refractive index in a complex direction~\cite{horsley2017}.
%
%
\begin{figure}[h!]
	\includegraphics[width=8cm]{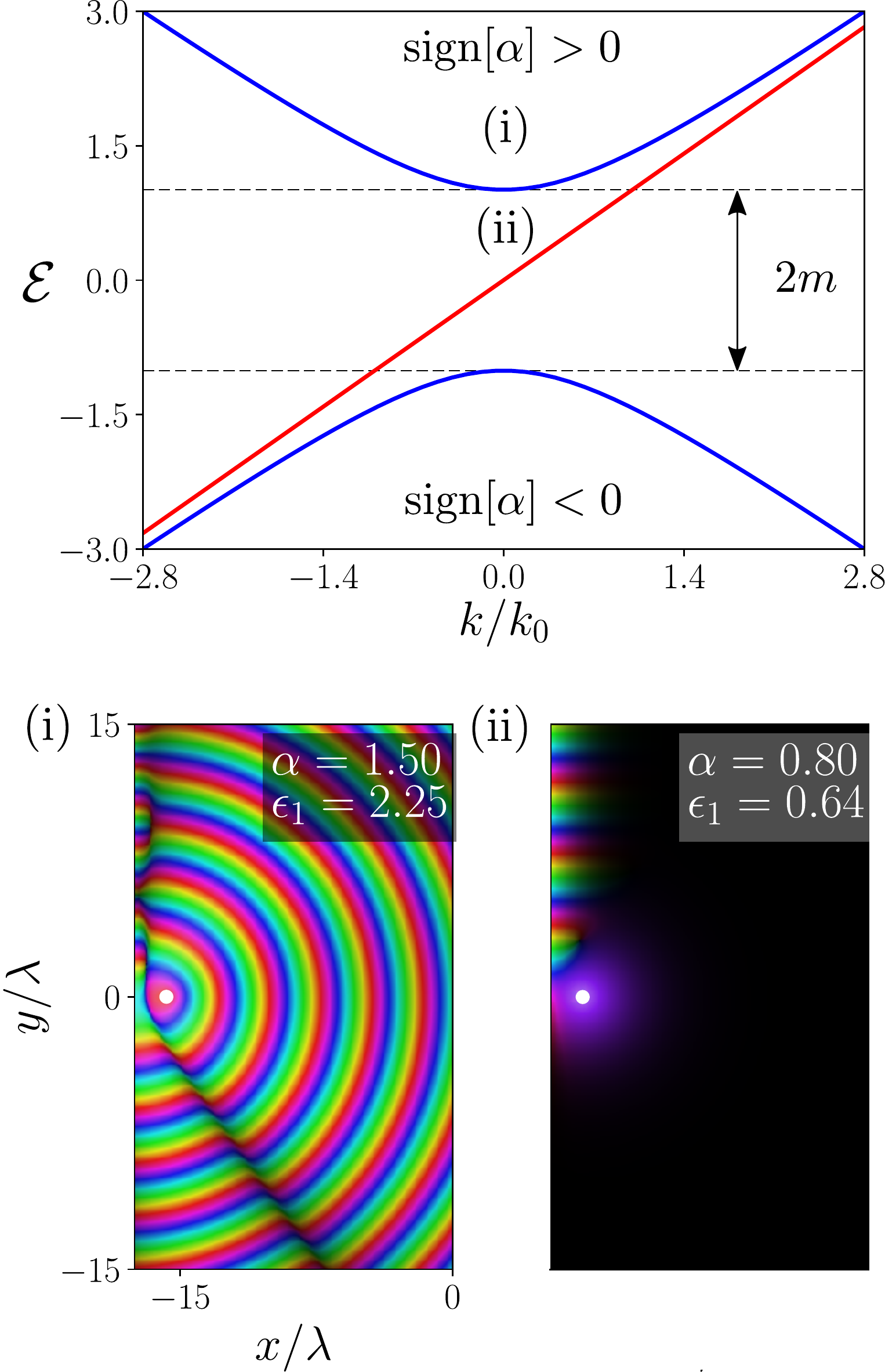}
	\caption{TM polarized electromagnetic waves in planar gyrotropic media obey a two dimensional Dirac equation, where the `energy' and `mass' are given by the material parameters $\mathcal{E}=\sqrt{\epsilon_{1}}$ and $m=\alpha/\sqrt{\epsilon_{1}}$ (see (\ref{eq:permittivity})).  (top) The value of the wave--vector (along say the $y$--axis) plotted as a function of $\sqrt{\epsilon_{1}}$ for fixed $\alpha=1$ with regions of bulk propagation in blue and the dispersion of the interface state in red (identical to the corresponding plot for the two dimensional Dirac equation in the presence of a domain wall).  (i--ii) Magnitude/phase maps (generated using COMSOL multiphysics) of the out of plane $H$ field for the two regions of parameter space indicated in the top panel (color indicates phase, brightness magnitude). The white dot shows the position of a point source used to generate the field, and a mirror is present along the line $x=-17.5\lambda$ ($\lambda=2\pi/k_0$).\label{fig:dirac_dispersion}}
\end{figure}
\par
Returning to the equations (\ref{eq:maxwell_complex}) we can see that the electric field components $E_{\pm}$ are not independent, and are related by
\[
	(\mathcal{E}-m)\frac{\partial E_{+}}{\partial z^{\star}}=-(\mathcal{E}+m)\frac{\partial E_{-}}{\partial z}
\]
Therefore we can reduce the system of three equations (\ref{eq:maxwell_complex}) to two equations.  For instance if we eliminate the $E_{-}$ field component we are left with
\begin{equation}
	\left(\begin{matrix}\mathcal{E}-m&2{\rm i}k_0^{-1}\frac{\partial}{\partial z}\\2{\rm i}k_0^{-1}\frac{\partial}{\partial z^{\star}}&\mathcal{E}+m\end{matrix}\right)\left(\begin{matrix}\psi_{a}\\\psi_{b}\end{matrix}\right)=0\label{eq:dirac_material1}.
\end{equation}
If we instead eliminate $E_{+}$ we obtain a very similar equation
\begin{equation}
\left(\begin{matrix}\mathcal{E}+m&-2{\rm i}k_0^{-1}\frac{\partial}{\partial z^{\star}}\\-2{\rm i}k_0^{-1}\frac{\partial}{\partial z}&\mathcal{E}-m\end{matrix}\right)\left(\begin{matrix}\psi_{a}\\\psi_{b}\end{matrix}\right)=0\label{eq:dirac_material2}.
\end{equation}
where $\psi_{a}=\sqrt{2}{\rm i}\mathcal{E}E_{\pm}$ (the positive sign in the subscript holds for equation (\ref{eq:dirac_material1}), the negative for (\ref{eq:dirac_material2})) and $\psi_{b}=\eta_0 H$.  A comparison with the Dirac equation for a fixed energy (\ref{eq:dirac2d}) shows that equation (\ref{eq:dirac_material1}) has the same form, in terms of dimensionless coordinates $k_0 z$, with the material parameters $\sqrt{\epsilon}_{1}$ and $\alpha/\sqrt{\epsilon_{1}}$ playing the role of energy and mass respectively.  Equation (\ref{eq:dirac_material2}) also has the same structure, but for the inverted coordinate system $x\to-x$ and inverted mass $m\to-m$.  Note that the definition of the permittivity tensor (\ref{eq:permittivity}) shows that changing the sign of the `energy' is equivalent to changing the sign of the gyrotropy $\alpha$ while keeping the sign of $\epsilon_{1}$ fixed, and one must use this interpretation to understand e.g. figure~\ref{fig:dirac_dispersion} .  With this identification of the material parameters with energy and mass, the `relativistic' dispersion relation is given by
\begin{equation}
	\left(\frac{\boldsymbol{k}}{k_0}\right)^{2}=\mathcal{E}^{2}-m^{2}=\frac{\epsilon_1^{2}-\alpha^{2}}{\epsilon_{1}}\label{eq:dispersion}
\end{equation}
which is the same expression as derived in~\cite{davoyan2013}, but now with a rather different interpretation (see figure~\ref{fig:dirac_dispersion}).  It is rather unexpected that the Dirac equation should be found lurking inside the optics of homogeneous gyrotropic media, in precisely the same way as it is found to be hidden inside the motion of non--relativistic particle within the honeycomb lattice~\cite{katsnelson2012}.
\par
\begin{figure}[h!]
\includegraphics[width=8.9cm]{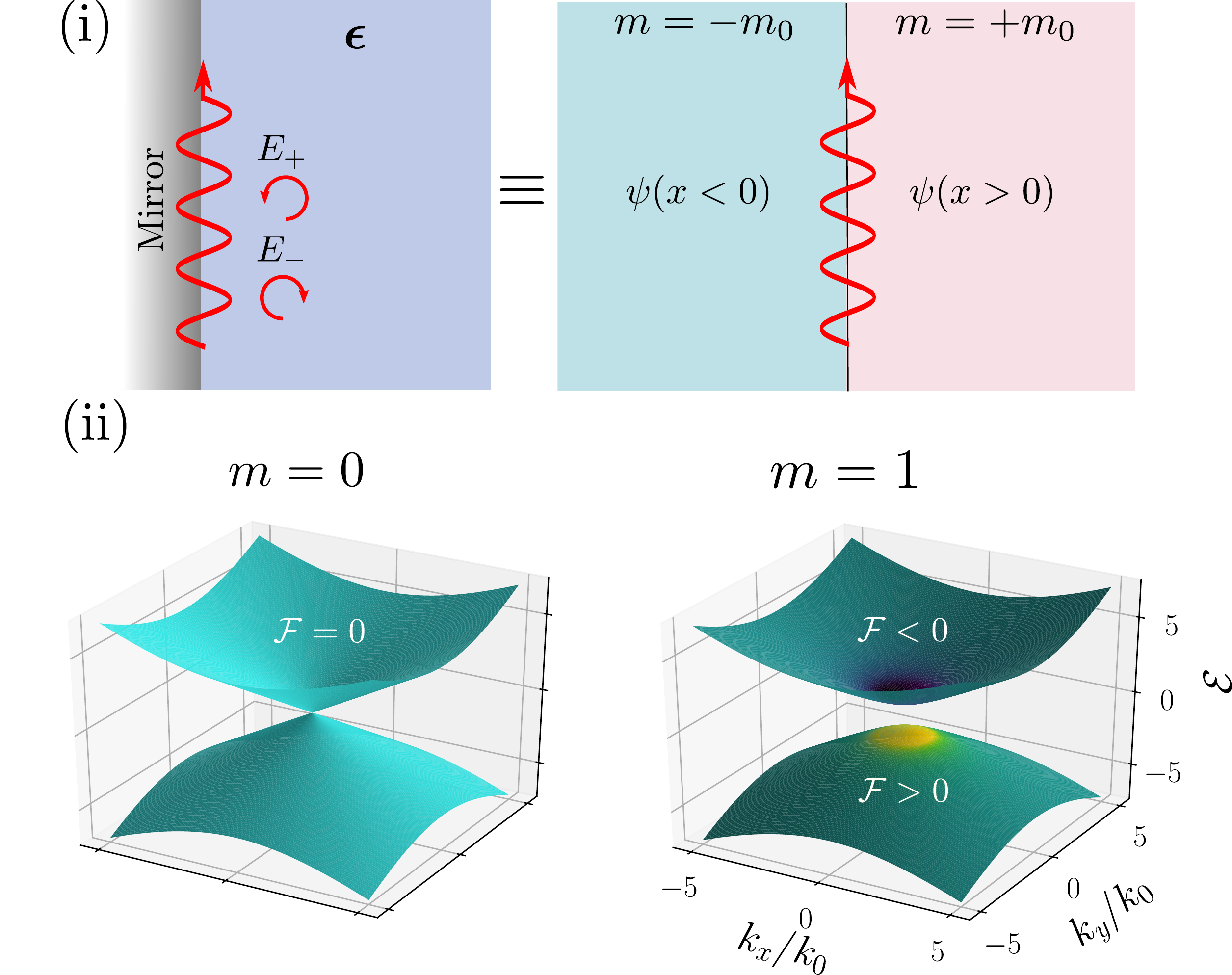}
	\caption{(i) An electromagnetic interface state propagating in a planar gyrotropic medium (permittivity tensor $\boldsymbol{\epsilon}$) and bound to the interface of a mirror at $x=0$ is governed by a 2D Dirac equation, where the mass changes sign at $x=0$.  The out of plane magnetic field combined with either of the two circulating electric field components $E_{\pm}$ determine the respective spinor $\psi$ in the two half spaces $x>0$ and $x<0$. (ii) Given the mapping onto the Dirac equation, the integral of the Berry curvature $\mathcal{F}$ in the space of parameters $\epsilon_{1},k_x,k_y$ determines the number of interface states.  The two plots show the dispersion (see figure~\ref{fig:dirac_dispersion}) with the coloring indicating the value of the Berry curvature (the integrand of (\ref{eq:berry_curvature})).\label{fig:berry_curvature}}
\end{figure}

This relationship to the Dirac equation is more than a formal trick.  Just as for Dirac particles, EM waves can be trapped at an interface where the `mass' $m=\alpha/\sqrt{\epsilon_{1}}$ changes sign (see~\cite{hasan2010} and the schematic in figure~\ref{fig:berry_curvature}i).  A neat way to illustrate this effect is to bound the medium with a perfect conductor.  If the conductor is in the $y$--$z$ plane, the boundary condition on the field is
\[
	E_{y}(x=0)=\frac{{\rm i}}{\sqrt{2}}\left[E_{+}(x=0)-E_{-}(x=0)\right]=0
\]
which means that the two Dirac equations (\ref{eq:dirac_material1}--\ref{eq:dirac_material2}) are coupled by the presence of the mirror, and both components of the two wave--functions are equal at $x=0$.  Given that the second of the two equations (\ref{eq:dirac_material2}) is equivalent to the first (\ref{eq:dirac_material1}) with $x\to-x$ and $\alpha\to-\alpha$ we see that electromagnetic wave propagation in such a gyrotropic medium bound by a perfect conductor maps directly onto the problem of the Dirac equation in two half spaces of opposite mass.  The major proviso is that this equivalence only holds for those modes of the Dirac equation where $\psi_{b}$ has mirror symmetry around $x=0$.  The edge state that occurs for the 2D Dirac equation between regions of oppositely signed mass~\cite{hasan2010} has this property and is thus equivalent to the mode that can exist at the interface between a mirror and a gyrotropic medium. A simple application of the boundary conditions to the solutions of (\ref{eq:dirac_material1}--\ref{eq:dirac_material2}) allows us to reclaim the result reported in~\cite{davoyan2013}, where this unidirectional mode is equal to
\begin{equation}
	\left(\begin{matrix}\psi_{a}\\\psi_{b}\end{matrix}\right)=\left(\begin{matrix}\sqrt{2}{\rm i}\mathcal{E}E_{+}\\\eta_0 H\end{matrix}\right)=\frac{1}{\sqrt{2}}\left(\begin{matrix}1\\{\rm i}\end{matrix}\right){\rm e}^{(-m k_0 x+{\rm i}\mathcal{E}k_0 y)}\label{eq:edge_state}
\end{equation}
but now we can recognize this as being a particular case of the well--known solution to the 2D Dirac equation~\cite{hasan2010}, with the wave--vector along the interface being that for a `massless' particle $k_y/k_0=\mathcal{E}$.  When the energy is less than the rest mass energy (i.e. the material is such that $\alpha^{2}>\epsilon_1$) then $H$ polarized waves cannot travel though the bulk of the medium, while the above solution (\ref{eq:edge_state}) persists at the interface, with only one possible direction of propagation (see lower panels of figure~\ref{fig:dirac_dispersion}).
%
%
\par
Although the simplicity of this system allows us to find this interface state (\ref{eq:edge_state}) directly, it is instructive to connect its existence with the known index theorems for Dirac operators in two dimensions~\cite{weinberg1981,shiozaki2012,volovik2003} (for spaces without a boundary, such index theorems are connected to the famous Atiyah--Singer index theorem~\cite{atiyah1963}).  To do this we adapt an argument due to Volovik~\cite{volovik2003} and define the spectral asymmetry $\nu(k_y)$ of the Dirac operator (with e.g. the position dependent mass $m(x)$ shown in figure~\ref{fig:berry_curvature}) as
\begin{equation}
	\nu(k_y)=\frac{1}{2}\sum_{n}\frac{\mathcal{E}_{n}}{|\mathcal{E}_{n}|}=\int_{-\infty}^{\infty}\frac{d\xi}{2\pi}{\rm Tr}[G(x,x)]\label{eq:spectral_asym}
\end{equation}
where we have introduced the Green function as $G=[\hat{H}_{\rm D}(k_y)+{\rm i}\xi)]^{-1}$, and `${\rm Tr}$' indicates a trace over matrix and spatial variables.  While the quantity $\nu$ typically represents the difference in the number of positive and negative \emph{energy} states with wave-vector $k_y$~\cite{shiozaki2012,volovik2003}, here the spectral asymmetry has a rather unusual meaning.  It tells us about the number of material parameters that are consistent with the wave having wave--vector component $k_y$ and frequency $k_0$.  For instance, for the case shown in the upper panel of figure~\ref{fig:dirac_dispersion} where $k>0$ we expect $\nu(k)$ to count one more mode (with value $1/2$) for positive values of $\mathcal{E}$ than for negative values, indicating that there is one more right-going mode available when the gyrotropy $\alpha$ is positive, compared to when it is negative.  But this does not yet tell us anything about whether the wave can propagate only one way.  The difference ($k_{\rm max}$ is some large positive value)
\begin{equation}
	N=\nu(k_{\rm max})-\nu(-k_{\rm max})\label{eq:index}
\end{equation}
counts the number of modes that start in the region of negative gyrotropy $\mathcal{E}<0$ and end with positive gyrotropy $\mathcal{E}>0$, as $k_y$ is increased.  For example, if this number equals $1$, it tells us that a single interface mode must move to cross the gap in the material parameter $\mathcal{E}$ shown in figure~\ref{fig:dirac_dispersion}.  This means that, for a fixed value of $k_0$ and a value of $\mathcal{E}=\sqrt{\epsilon_{1}}$ in the gap, we must have an odd number of interface states (for a linear dependence on $\mathcal{E}$ this number is one), and hence an asymmetry in the direction of power transported at the edge.  This is a conceptually different way to predict the number of interface states in continuous media, and does not require us to discuss the dispersion of the material parameters.
\par
A practical problem with using (\ref{eq:spectral_asym}) is that we don't in general have an expression for the Green function $G$.  However, because $k_{\rm max}$ is assumed large, the Green function will be exponentially localized, and we can use a semi--classical approximation (replacing the step change in $m(x)$ with a smooth function).  As shown in appendix~\ref{ap:index}, using an argument of multiple scales, the Green function can be written to first order as $G=\mathcal{G}-\frac{{\rm i}}{2}\mathcal{G}\left\{H,\mathcal{G}\right\}$, where $\{,\}$ is a Poisson bracket with respect to $x$ and $k_x$.  The semi--classical Green function $\mathcal{G}$ appearing in this formula is given by
\begin{equation}
	\mathcal{G}(x,x,k_y)=\int_{-\infty}^{\infty}\frac{d k_x}{2\pi}\frac{1}{H_{\rm D}(x,k_x,k_y)+{\rm i}\xi}\label{eq:semi-classical-gf}
\end{equation}
where $H_{\rm D}$ is the `classical' Dirac Hamiltonian (obtained through the replacement $-{\rm i}\partial_{x}\to k_x$).  Applying this semiclassical approximation, the number $N$ is given by the integral of a three form
\begin{equation}
	N=-\frac{1}{24\pi^{2}}e^{\mu\nu\sigma\tau}{\rm tr}\int dS_{\mu}(\mathcal{G}\partial_{\nu}\mathcal{G}^{-1})(\mathcal{G}\partial_{\sigma}\mathcal{G}^{-1})(\mathcal{G}\partial_{\tau}\mathcal{G}^{-1})\label{eq:number_of_modes_mt}
\end{equation}
where the indices refer to the coordinates $(\xi,x,k_x,k_y)=(\zeta^{0},\zeta^{1},\zeta^{2},\zeta^{3})$ and the integral is over a closed surface at infinity (note, one must be careful to ensure $\mathcal{G}$ vanishes at infinity).  The expression (\ref{eq:number_of_modes_mt}) is well known in quantum field theory, where it appears as the divergence of the Chern--Simons current~\cite{srednicki2007}.  When $\mathcal{G}$ is an element of SU(2), the above integral represents the winding number of the mapping from the closed hypersurface in $\zeta_{\mu}$ space to the three sphere representation of SU(2).  Using the fact that the integrand is locally an exact form, the integral (\ref{eq:number_of_modes_mt}) can be transformed into a pair of integrals over hypersurfaces of constant $x$
\begin{equation}
	N=\nu(x_{\rm max})-\nu(-x_{\rm max})
\end{equation}
which for the Dirac Hamiltonian equals (see appendix~\ref{ap:index})
\begin{align}
	\nu(x\lessgtr 0)&=\frac{1}{4\pi}\int d^{2}\boldsymbol{k}\;\boldsymbol{n}\cdot\left(\frac{\partial\boldsymbol{n}}{\partial k_x}\times\frac{\partial\boldsymbol{n}}{\partial k_y}\right)\nonumber\\
    &=\frac{1}{2}{\rm sign}[m(x\lessgtr 0)]\label{eq:winding}
\end{align}
with $\boldsymbol{n}=(\boldsymbol{k}/k_0+m(x)\hat{\boldsymbol{z}})/\sqrt{\boldsymbol{k}^{2}/k_0^{2}+m(x)^{2}}$.  Therefore, when $\mathcal{E}<m$, only a single mode exists at the interface between a mirror and a gyrotropic medium (configured as shown in figure~\ref{fig:berry_curvature}), with ${\rm sign}[k_y]={\rm sign}[\alpha]$, in agreement with the dispersion shown in figure~\ref{fig:dirac_dispersion}, and the exact solution (\ref{eq:edge_state}).
\par
This winding number can be related to the wave--functions of the system.  We write the real vector $\boldsymbol{n}$ as the equivalent point on the Bloch sphere
\begin{align}
	\boldsymbol{n}&\to\left(\begin{matrix}\cos(\theta/2)\\\sin(\theta/2){\rm e}^{{\rm i}\phi}\end{matrix}\right)\nonumber\\
    &=\frac{1}{\sqrt{2\mathcal{E}}}\left(\begin{matrix}\sqrt{\mathcal{E}+m}\\\sqrt{\mathcal{E}-m}{\rm e}^{{\rm i}\phi_{k}}\end{matrix}\right)=\left(\begin{matrix}\psi_{a}\\\psi_{b}\end{matrix}\right)\label{eq:equivalent-spinor}
\end{align}
where $\boldsymbol{k}=|\boldsymbol{k}|(\cos(\phi_k)\hat{\boldsymbol{x}}+\sin(\phi_k)\hat{\boldsymbol{y}})$, $\boldsymbol{n}=\cos(\theta)\hat{\boldsymbol{z}}+\sin(\theta)(\cos(\phi)\hat{\boldsymbol{x}}+\sin(\phi)\hat{\boldsymbol{y}})$, and we used (\ref{eq:dispersion}) along with the definition of $\boldsymbol{n}$ in terms of $\boldsymbol{k}$ and $m$.  This spinor representation of the unit vector $\boldsymbol{n}$ is simply the solution to our Dirac equation (\ref{eq:dirac_material1}), with the sign of $\mathcal{E}$ determining whether the point is in the upper or lower hemisphere.  Translating (\ref{eq:winding}) into this representation of $\boldsymbol{n}$, we find
\begin{equation}
	\nu=\frac{1}{2\pi}\int\mathcal{F}_{-}d^{2}\boldsymbol{k}=-\frac{{\rm i}}{2\pi}\int d^{2}\boldsymbol{k}\boldsymbol{\nabla}_{\boldsymbol{k}}\psi^{\dagger}_{-}\times\boldsymbol{\nabla}_{\boldsymbol{k}}\psi_{-}\cdot\hat{\boldsymbol{z}}\label{eq:berry_curvature}
\end{equation}
This is the integral of the Berry curvature~\cite{nakahara2003,berry1984}.  Notice that, for a fixed sign of $m$ the unit vector $\boldsymbol{n}$ only explores half of the sphere (hence the factor of $1/2$ in (\ref{eq:winding})).  The subscript `$-$' in (\ref{eq:berry_curvature}) indicates that the negative sign of $\mathcal{E}$ is taken (see figure~\ref{fig:berry_curvature}), which means that we correspondingly explore only half of the Bloch sphere.  Because the net Berry curvature of both bands is zero, note that (\ref{eq:berry_curvature}) can be written in terms of either $\mathcal{F}_{-}$ or $\mathcal{F}_{+}$.  As illustrated in panel ii of figure~\ref{fig:berry_curvature}, this curvature arises from the separation of the dispersion relation into two bands, as a function of the material parameter $\mathcal{E}=\sqrt{\epsilon_{1}}$.  The curvature is concentrated around the points where $\mathcal{E}=\pm m$ (or equivalently $\epsilon_{1}=\pm\alpha$).  These points have a special significance in optics, being where ${\rm det}[\boldsymbol{\epsilon}]=0$, and indicating a zero in the refractive index for propagation in a complex direction~\cite{horsley2017}, which are the points where the wave becomes an analytic function of position, and is thus forced to propagate in only one direction.
\par
Although this calculation appears superficially identical to some well established results, what we have done is rather different from much of the existing work on this topic e.g.~\cite{haldane2008,lu2014,silveirinha2015}.  The question we're asking is not ``given these fixed material parameters with this dependence on frequency, are there unidirectional interface states present in a band gap?''.  It is rather ``If I vary the material parameter $\sqrt{\epsilon_{1}}$ at a \emph{fixed} frequency, are there unidirectional interface states present when $\sqrt{\epsilon_1}$ takes values where bulk propagation is forbidden?''.  This method does not require one to provide the dispersion of the material parameters, and indicates a more general method for applying the mathematics of topology to optics.
%
%
\begin{figure}[h!]
\includegraphics[width=8cm]{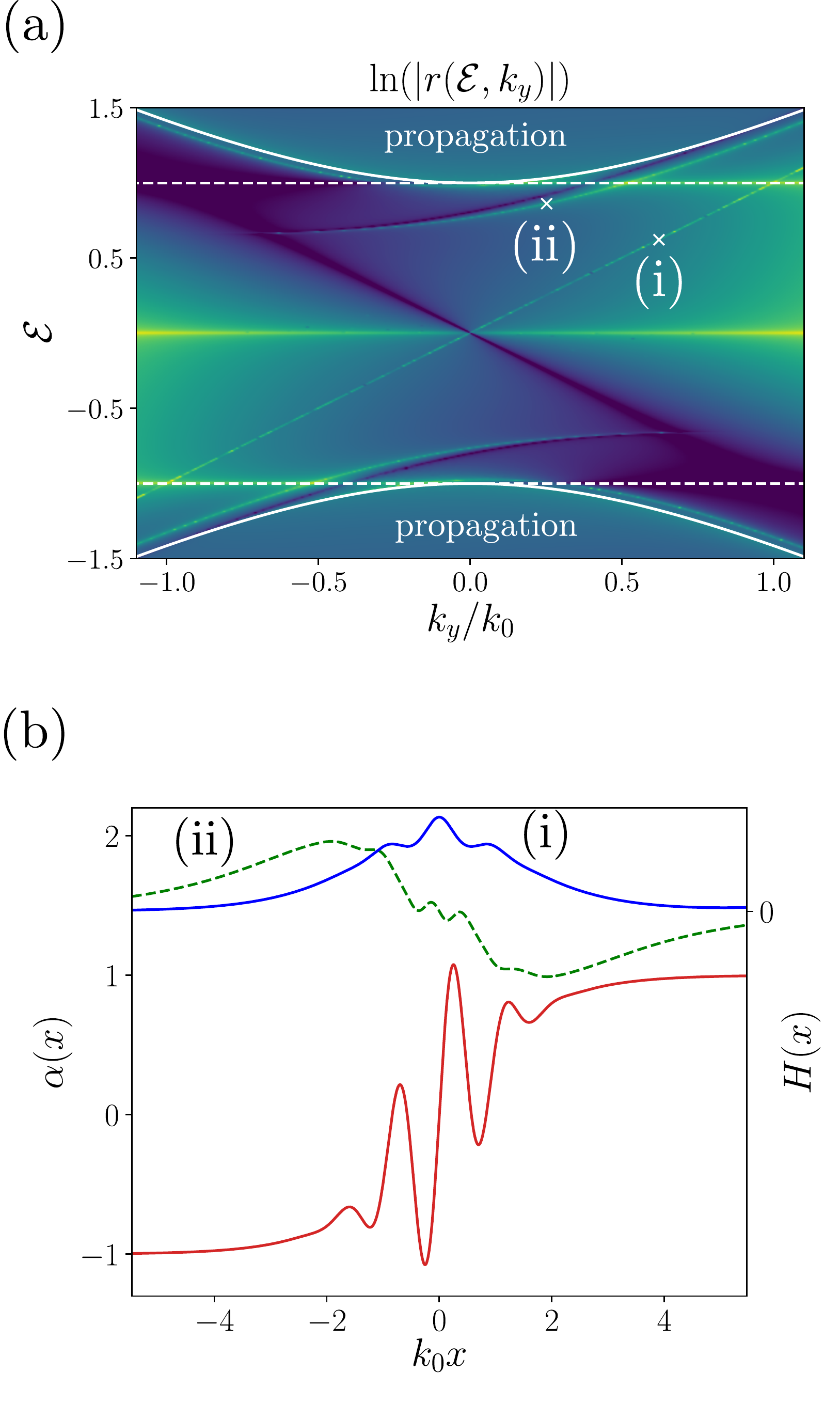}
\caption{(a) Logarithm of reflection coefficient $\ln(|r(\mathcal{E},k_y)|)$ for the medium with properties given by (\ref{eq:topological_material}) and polarization $E_z=-\tan(\theta)\eta_0 H_z$, verifying the prediction of (\ref{eq:mode_number}) with two modes moving from $k_y,\mathcal{E}<0$, to $k_y,\mathcal{E}>0$ (another two such modes are evident in the orthogonal polarization). Between the two horizontal dashed white lines no propagation is allowed.  (b) The functional form of $\alpha(x)$ which has been chosen as an arbitrary function that changes sign around $x=0$ (the function $\alpha'$ plays no role for this polarization, but is also assumed to also change sign).  (i--ii) show the functional form of the two interface modes, with the corresponding parameters shown in panel (a)  \label{fig:reflection_figure}}
\end{figure}
\par
So far we have only treated homogeneous gyrotropic media, for a single polarization.  This was mainly for the purposes of simplicity.  There is no fundamental restriction on the applicability of (\ref{eq:number_of_modes_mt}); so long as the Green function is suitably defined, it may be use to calculate the number of modes crossing any region of parameter space where propagation is not allowed.  As we shall see, this includes any inhomogeneous electromagnetic medium.  To generalize the above analysis we first recognise that Maxwell's equations in any medium can be written in the same form as the Dirac equation for a six component wave function $\psi$ (see appendix~\ref{ap:chern_maxwell})
\begin{equation}
	\left[-{\rm i}k_0^{-1}\gamma_{i}\partial_{i}-\chi'\right]\psi=\hat{H}\psi=\mathcal{E}\psi.\label{eq:general_dirac}
\end{equation}
where $\psi=(\boldsymbol{E},\eta_0\boldsymbol{H})^{\rm T}$, and $\chi'$ is the traceless part of the material tensor
\begin{equation}
	\chi=\left(\begin{matrix}\boldsymbol{\epsilon}&\boldsymbol{\xi}\\\boldsymbol{\xi}^{\dagger}&\boldsymbol{\mu}\end{matrix}\right)=1_{6}\mathcal{E}+\chi'\label{eq:material_tensor}
\end{equation}
where $\boldsymbol{\epsilon}$, $\boldsymbol{\mu}$ and $\boldsymbol{\xi}$ are the permittivity, permeability and bianisotropy respectively.  We have also introduced the analogues of Dirac's $\gamma$ matrices 
\begin{equation}
	\gamma_{i}=\left(\begin{matrix}\boldsymbol{0}&-\boldsymbol{L}_{i}\\\boldsymbol{L}_{i}&\boldsymbol{0}\end{matrix}\right)
\end{equation}
where $\boldsymbol{L}_{i}$ are a $3\times3$ representation of the spin 1 matrices~\cite{barnett2014}.  In passing we note that in this way of writing Maxwell's equations $\chi'$ can be understood as contributing effective gauge field and mass terms to this Dirac equation (something which requires a little unpacking in the standard formalism~\cite{liu2015}). The only wrinkle with our analogy is that the $\gamma_{i}$ no longer constitute a Clifford algebra (the square of $\gamma_{i}\partial_{i}$ is not the Laplacian $\boldsymbol{\nabla}^{2}$, but rather the transverse Laplacian $\boldsymbol{\nabla}\times\boldsymbol{\nabla}\times$), although this does not affect our results.  As in the case of the gyrotropic medium bound by a mirror, the `energy' in this Dirac equation is a material parameter.  Here it is $1/6$ of the trace of the material tensor (\ref{eq:material_tensor}).  In appendix~\ref{ap:chern_maxwell} we show that if we assume the `classical' Hamiltonian appearing in (\ref{eq:semi-classical-gf}) has `chiral' symmetry (c.f.~\cite{aoki2014,shiozaki2012})
\begin{equation}
	THT^{\dagger}=-H\label{eq:chiral_symmetry}
\end{equation}
then the consequent spectral symmetry of the `classical' Hamiltonian means that the number of modes that cross from $\mathcal{E}<0$ to $\mathcal{E}>0$ with increasing $k_y$ is again given by formula (\ref{eq:number_of_modes_mt}) (without this symmetry there is an additional term in the spectral asymmetry, which we do not want to consider here).  In this general case there is no simple formula such as (\ref{eq:winding}), but an integration over $\zeta_0$ in (\ref{eq:number_of_modes_mt}) leads to
\begin{multline}
	N=\frac{{\rm i}}{2\pi k_0^{2}}\left.e^{pq}\sum_{\substack{\mathcal{E}_n> 0\\\mathcal{E}_m<0}}\int d^{2}\boldsymbol{k}\frac{\langle m|\gamma_{p}|n\rangle\langle n|\gamma_{q}|m\rangle}{\left(\mathcal{E}_{m}-\mathcal{E}_{n}\right)^{2}}\right|_{x=x_{\rm max}}\\
    -(x_{\rm max}\to-x_{\rm max})\label{eq:berry_2}
\end{multline}
where $e^{pq}$ is the antisymmetric unit tensor with indices $p,q\in\{1,2\}$.  Equation (\ref{eq:berry_2}) is again the difference in the integrated Berry curvature between the two positions $x=\pm x_{\rm max}$, and again it is written in terms of the possible material parameters $\mathcal{E}_{n}$.  Each of the two integrals of the Berry curvature (\ref{eq:berry_2}) are known as Chern numbers~\cite{nakahara2003}.  The form of the Hamiltonian (\ref{eq:general_dirac}) leads to the matrix elements of the angular momentum operators $\gamma_{i}$ appearing in (\ref{eq:berry_2}): it is only modes where both angular momenta have a non--zero complex overlap $\langle m|\gamma_{p}|n\rangle$ that contribute to the Berry curvature.  As an example application of formula (\ref{eq:berry_2}) we consider the following traceless material tensor
\begin{widetext}
\begin{equation}
	\chi'=\left(\begin{matrix}-{\rm i}\left(\alpha\cos^{2}(\phi)+\alpha'\sin^{2}(\phi)\right)\boldsymbol{L}_{3}&-{\rm i}(\alpha-\alpha')\sin(\phi)\cos(\phi)\boldsymbol{L}_{3}\\-{\rm i}(\alpha-\alpha')\sin(\phi)\cos(\phi)\boldsymbol{L}_{3}&-{\rm i}\left(\alpha'\cos^{2}(\phi)+\alpha\sin^{2}(\phi)\right)\boldsymbol{L}_{3}\end{matrix}\right)\label{eq:topological_material}
\end{equation}
\end{widetext}
where $\phi\in[-\pi,\pi]$.  This is a medium exhibiting both gyrotropy and bianisotropy (see supplementary material for the genesis of expression (\ref{eq:topological_material}).  A calculation of the number of interface states existing in an inhomogeneous region where $\alpha$ and $\alpha'$ change with position, via formula (\ref{eq:berry_2}) yields
\begin{multline}
	N={\rm sign}[\alpha(x_{\rm max})]+{\rm sign}[\alpha'(x_{\rm max})]\\
    -{\rm sign}[\alpha(-x_{\rm max})]-{\rm sign}[\alpha'(-x_{\rm max})]\label{eq:mode_number}
\end{multline}
which implies that a maximum of four interface states cross from $k_{y},\mathcal{E}<0$ to $k_y,\mathcal{E}>0$, when both $\alpha$ and $\alpha'$ change sign over the interval $x\in[-x_{\rm max},x_{\rm max}]$.  Figure~\ref{fig:reflection_figure}a shows the reflection coefficient for one polarization incident onto such a medium, with two of the four interface modes evidently crossing the region where propagation is not allowed (details of the calculation are given in appendix~\ref{ap:example}).  Panel b of this figure shows the functional form of these interface modes, with one having even parity and the other odd parity.
\par
Finally, to make contact with existing work on this topic, we show that the above formalism can also reproduce the known results for periodic media.  To do this we consider the eigenvalue problem (\ref{eq:general_dirac}), written for frequency (rather than the trace of $\chi$)
\begin{equation}
    \left(-{\rm i}\chi^{-1}\gamma_j\partial_j-k_g\right)|\psi\rangle=\hat{H}|\psi=\mathcal{E}|\psi\rangle
\end{equation}
where the frequency has been written as $k_0=\mathcal{E}+\omega_g/c$, with $\omega_g$ the frequency in the middle of the band gap of interest.  It is imagined that the medium of interest is periodic in $x$ and $y$ as $x\to\pm\infty$, but the unit cell is different on the two sides.  We imagine the unit cell changes slowly as a function of $x$, compared to its size.  To describe this we split the $x$ dependence of the Hamiltonian into two parts $\hat{H}(-{\rm i}\partial_x,x)\to\hat{H}(-{\rm i}\partial_x,x;x)$, where the Hamiltonian is periodic in the first argument, and the second argument describes the change in this periodicity.  Because the Green function $G(x,x')$ is exponentially localized, it can be expanded around the Green function for a periodic medium $\mathcal{G}(x,x';x_1)$ with the periodicity at the average position $x_1=(x+x')/2$.  To leading order this expansion is given by the analogue of (\ref{eq:semi-classical-gf})
\begin{equation}
    G\sim\mathcal{G}-\frac{\rm i}{2}\mathcal{G}\left\{H,\mathcal{G}\right\}
\end{equation}
The Poisson bracket is here taken with respect to the Bloch vector $K_x$ and the average position $x_1$, and the product of matrices must be understood as a summation over both spatial and matrix indices. 
As above in equation (\ref{eq:index}) we can calculate the difference in the spectral asymmetry $\nu(x_1)$ between $x_1=-\infty$ and $x_1=+\infty$, which will tell us how many modes cross the gap as we move through the medium.  Exactly the same procedure can be followed as that leading to equation (\ref{eq:berry_2}), and the number of interface states will thus be given by the difference in the total Berry curvature for the bands $\omega<\omega_g$, between the two periodic media at $x\to\pm\infty$.
\par
In this work we have examined an analogy between the Dirac equation---which describes particles of half integer spin---and Maxwell's equation in material media.  The purpose was to clarify and extend the application of topology in optics, a relationship that is well developed for Dirac--type Hamiltonians.  Starting from a theory due to Volovik~\cite{volovik2003}, we found a quite general formula (\ref{eq:winding}), where the number of interface states between adjoining electromagnetic media equals the integrated divergence of the Chern--Simons current~\cite{srednicki2007}.  When applied to planar interfaces, this formula reduces to a sum over integrals of the Berry curvature over $\boldsymbol{k}$ space, commonly known as Chern numbers.  Unlike many previous results in the context of optics, the version of the theory presented here is not restricted to periodic media (although it can be applied to such media, as shown above), and does not require one to know how the material parameters disperse with frequency.   We have given an example showing that the integrated Berry curvature can be used to predict the number of interface states at a fixed frequency, through considering the dependence of the electromagnetic modes on the material parameters.  It is anticipated that an analogous theory could also be applied to non--Hermitian electromagnetic materials, which will be the topic of future work.

\acknowledgements
The author acknowledges very useful conversations with Tom Philbin, and Andrey Shytov for his enlightening lectures on topology.  Financial support was provided by the Royal Society and TATA (RPG-2016-186).

\appendix
\section{Counting modes of the Dirac operator\label{ap:index}}
\par
As shown in the main text, the propagation of electromagnetic waves in a gyrotropic medium bound by a mirror can be described using the Dirac equation in two half spaces with mass of the opposite sign, connected at the mirror (see figure~\ref{fig:mode_counting}i and figure 2i in the main text).  We now adapt the argument of Volovik~\cite{volovik2003} to show the relationship between the number of states bound to the mirror interface, and a winding number computed from the values of $\boldsymbol{k}$ and $m(x)$, associated with propagation in the bulk of the two gyrotropic media.  This simple example shows that writing Maxwell's equations in the form of a Dirac operator allows us to establish a rigorous link between the number of interface states and the mathematics of topology.  We show the generalization of this to the full Maxwell equations in the next section.
\par
We calculate an integer $N$, which is the difference between the number of modes that have end points $\mathcal{E}(k_y\to-\infty)<0$, $\mathcal{E}(k_y\to+\infty)>0$, and those with end points $\mathcal{E}(k_y\to-\infty)>0$, $\mathcal{E}(k_y\to+\infty)<0$ (see schematic in figure~\ref{fig:mode_counting}ii). This integer can be written as
\begin{equation}
	N=\nu(k_0,k_{y}\to+\infty)-\nu(k_0,k_{y}\to-\infty)\label{eq:number_of_modes}
\end{equation}
where $\nu$ is the so--called `spectral asymmetry'~\cite{shiozaki2012}, defined as
\begin{align}
	\nu(k_0,k_y)&=\frac{1}{2}\sum_{n}{\rm sign}[\mathcal{E}_{n}(k_0,k_y)]\nonumber\\
    &={\rm Tr}\sum_{n}\int \frac{d\xi}{2\pi}\frac{|n\rangle\langle n|}{\mathcal{E}_{n}(k_0,k_y)+{\rm i}\xi}\nonumber\\
    &={\rm tr}\int dx\int \frac{d\xi}{2\pi}G(x,x)\label{eq:index_thm}
\end{align}
where, as in the main text `$\rm Tr$' indicates a sum over both spatial and matrix indices so that
\[
    {\rm Tr}\,|n\rangle\langle n|=1
\]
In our case the eigenvalues $\mathcal{E}_{n}$ represent e.g. the material parameter $\sqrt{\epsilon_{1}}$, ranging over all possible values that support a wave of fixed $k_0$ and $k_y$ (i.e. we ask what possible materials could support a wave with these values of $k_0$ and $k_y$).  In our definition of $\nu$ we have introduced $|n\rangle$ as the eigenfunctions of our Dirac Hamiltonian
\begin{multline}
	\left[-{\rm i}k_0^{-1}\left(\sigma_{x}\frac{\partial}{\partial x}+{\rm i}k_{y}\sigma_{y}\right)+m(x)\sigma_{z}\right]|n\rangle\\
    =\hat{H}|n\rangle=\mathcal{E}_{n}|n\rangle\label{eq:dirac_2}
\end{multline}
and the Green function as the solution to
\begin{equation}
	\left[\hat{H}\left(-{\rm i}\frac{\partial}{\partial x},k_y,x\right)+{\rm i}\xi\right]G(x,x',k_y)=\delta(x-x').\label{eq:green_equation}
\end{equation}
As illustrated in figure~\ref{fig:mode_counting}ii, to contribute to (\ref{eq:number_of_modes}) a mode must cross the line $\mathcal{E}=\sqrt{\epsilon_{1}}=0$, which is a region of parameter space where there can be no propagation in the bulk of either medium.  We are thus calculating the difference between the number of up and down--going waves bound to the interface at $x=0$, for the range of material parameters where propagation is not allowed.
\par
Until now we have treated the modes as a discrete set, when in fact they are part of a continuum.  It is usually the case that physical processes do not care about such sleights of hand.  But here there are some very delicate issues regarding what happens at the boundary of the system, which are not explicitly dealt with in~\cite{volovik2003}.  In order to make everything well defined we impose the boundary condition that the field tends to zero at infinity.  To do this in a concrete way we assume that the `mass' $m(x)$ remains constant for a large distance either side of the interface at $x=0$ before slowly but continuously increasing to $\pm\infty$ without changing sign (see figure~\ref{fig:mode_counting}i).  This causes all the modes to decay to zero at infinity and makes the spectrum discrete.  It does not however affect the modes of interest, which are bound to the interface at $x=0$. 
\begin{widetext}
\onecolumngrid
\begin{figure}[h]
	\begin{center}
	\includegraphics[width=16cm]{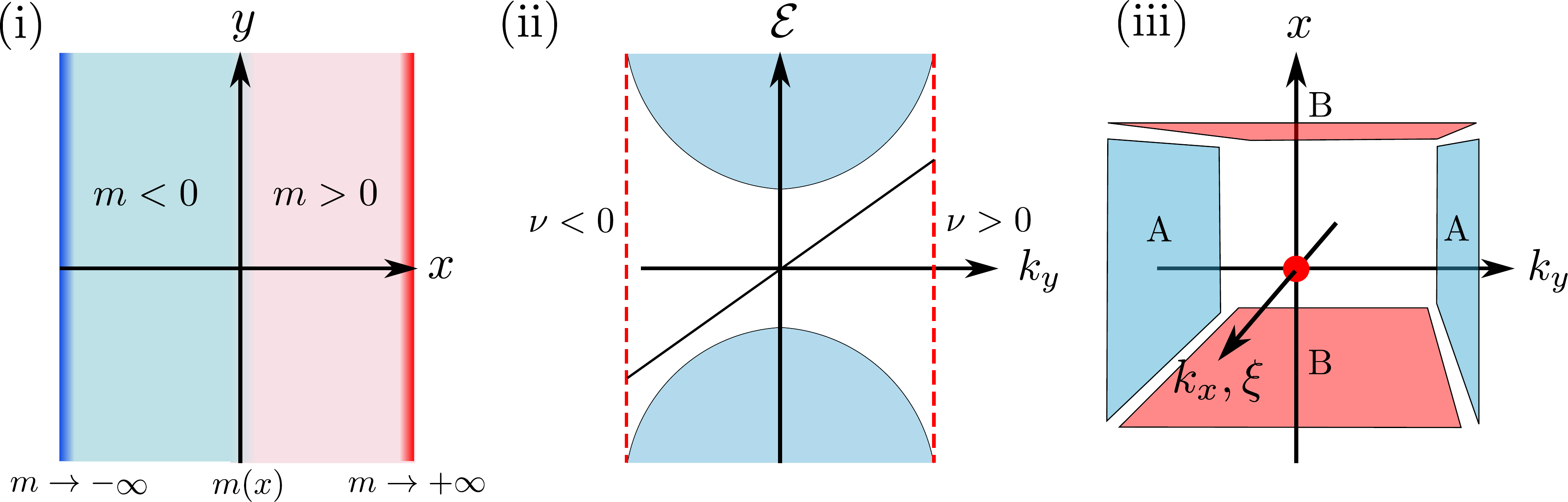}
    \caption{(i) We consider the electromagnetic states bound to the interface between a mirror and a gyrotropic medium as equivalent to the modes of the 2D Dirac equation bound to the interface between regions where the mass changes sign. We approximate this interface as with a region where $m$ smoothly changes sign.  In order to make the spectrum of our Dirac operator discrete, we also assume that the mass becomes infinite as $x\to\infty$ (without changing sign along the way). (ii) The system shown in panel i supports propagating waves for $\mathcal{E}^{2}>m^{2}$ (i.e. when $\epsilon_{1}^{2}>\alpha^{2}$), shown as the blue shaded regions, with an interface state connecting these two regions.  To work out the number of interface states that connect these regions in parameter space, we count the number of modes that cross the line $\mathcal{E}=\sqrt{\epsilon_{1}}=0$, where no propagation is allowed.  This counting is done in terms of the difference in spectral asymmetry $\nu(k_y)$ (defined in (\ref{eq:index_thm})) at $k\to+\infty$ and $k\to-\infty$, shown schematically by the vertical dashed lines.  (iii) The number of edge states $N$ is given as the difference between two integrals over hypersurfaces of constant $k_y$ (labeled A).  Using Stokes's theorem this can be transformed into integrals over constant $x$ (labeled B).\label{fig:mode_counting}}
    \end{center}
\end{figure}
\twocolumngrid
\end{widetext}
\par
To make progress we also assume that the step--like change in the sign of $m(x)$ across $x=0$ is the limit of a smooth function.  Importantly, our mode counting (\ref{eq:number_of_modes}) requires and evaluation of the Green function at large values of $k_y$.  In the limit $k_y\to\pm\infty$, the Green function becomes exponentially localized $G(x,x',k_y)\sim {\rm e}^{-|k_y||x-x'|}$.  Given this localization, we can apply a semi--classical expansion of $G(x,x')$, where $m(x)$ is assumed to vary slowly over the region where $G$ is different from zero.  This semiclassical expansion is performed through adopting the coordinates $x_{1}=(x+x')/2$ and $x_{2}=x-x'$, assuming the Hamiltonian varies slowly with respect to $x_1$.  In terms of these coordinates the Hamiltonian approximates to
\begin{align}
        \hat{H}\left(-{\rm i}\partial_x,k_y,x\right)&=\hat{H}\left(-{\rm i}\left(\partial_{x_2}+\frac{1}{2}\partial_{x_{1}}\right),k_y,x_1+\frac{1}{2}x_2\right)\nonumber\\
        &\sim\hat{H}(-{\rm i}\partial_{x_2},k_y,x_1)-\frac{{\rm i}}{2}\frac{\partial\hat{H}}{\partial k_x}\frac{\partial}{\partial x_1}\\
        &-\frac{{\rm i}}{4}\frac{\partial^2\hat{H}}{\partial x_1\partial k_x}+\frac{1}{4}\left(\frac{\partial\hat{H}}{\partial x_1}x_2+x_2\frac{\partial\hat{H}}{\partial x_1}\right)\label{eq:ham_grad_exp}
\end{align}
where the derivative of the Hamiltonian with respect to $k_x$ means a derivative with respect to the first (operator) argument.  Writing the Green function as $G(x_1,x_2,k_y)=(2\pi)^{-1}\int_{-\infty}^{\infty} dk_x G(x_1,k_x,k_y){\rm e}^{{\rm i}k_x x_2}$ and applying (\ref{eq:ham_grad_exp}) to (\ref{eq:green_equation}), after an integration by parts we find
\begin{multline}
	\bigg[H\left(k_x,k_y,x_{1}\right)+{\rm i}\xi\\
    +\frac{{\rm i}}{2}\left(\frac{\partial H}{\partial x_1}\frac{\partial}{\partial k_x}-\frac{\partial H}{\partial k_x}\frac{\partial}{\partial x_{1}}\right)\bigg]G(x_{1},k_x,k_y)=1
\end{multline}
where $H$ is now the classical Hamiltonian obtained by the replacements $-{\rm i}\partial/\partial x\to k_x$ and $x\to x_1$.  To leading order the Green function is then given by
\begin{equation}
	G(x_{1},k_x,k_y)\sim\mathcal{G}-\frac{{\rm i}}{2}\mathcal{G}\left\{H,\mathcal{G}\right\}\label{eq:green_wkb}
\end{equation}
where $\{H,\mathcal{G}\}=\partial_{x}H\partial_{k_x}\mathcal{G}-\partial_{k_x}H\partial_{x}\mathcal{G}$ is the Poisson bracket and,
\begin{equation}
	\mathcal{G}=\frac{1}{H+{\rm i}\xi}.
\end{equation}
Substituting (\ref{eq:green_wkb}) into (\ref{eq:index_thm}) and applying (\ref{eq:number_of_modes}) we obtain the following formula for the spectral asymmetry
\begin{equation}
	\nu(k_y)=-\frac{{\rm i}}{2}{\rm tr}\int \frac{d\xi}{2\pi}\int dx\int\frac{d k_x}{2\pi}\left(\mathcal{G}\left\{H,\mathcal{G}\right\}\right)
\end{equation}
The classical Hamiltonian has a `chiral symmetry' where the eigenvalues come in pairs of opposite sign equal to $\pm\sqrt{\boldsymbol{k}^{2}/k_0^2 + m^2}$, and thus the zeroth order term in (\ref{eq:green_wkb}) cannot contribute to the spectral asymmetry $\nu(k_y)$.  Introducing the coordinates $(\zeta_0,\zeta_1,\zeta_2,\zeta_3)=(\xi,x,k_x,k_y)$, and applying the identity $\partial \mathcal{G}=-\mathcal{G}\partial\mathcal{G}^{-1}\mathcal{G}$ the above integral can be written as a three form
\begin{equation}
	\nu(k_y)=\frac{1}{24\pi^{2}}{\rm tr}\int d^{3}\zeta e^{ijk}\big(\mathcal{G}\partial_{i}\mathcal{G}^{-1}\mathcal{G}\partial_{j}\mathcal{G}^{-1}\mathcal{G}\partial_{k}\mathcal{G}^{-1}\big)\label{eq:index_integral}
\end{equation}
where $d^{3}\zeta=d\zeta_{0}d\zeta_{1}d\zeta_{2}$, the indices $i,j,k$ run over $0,1,2$, and we used the cyclic property of the trace .  Because our boundary conditions ensure that $\mathcal{G}$ tends to zero when any of the $\zeta_{\mu}$ tend to infinity, the difference in $\nu$ at two arbitrarily large but finite values $k_y=k_{\rm max}$ can be written as an integral over a closed hyper--surface $\partial V$
\begin{multline}
	\nu(k_{\rm max})-\nu(-k_{\rm max})=\\
    -\frac{1}{24\pi^{2}}{\rm tr}\int_{\partial V} dS_{\mu}e^{\mu\nu\sigma\tau}\left(\mathcal{G}\partial_{\nu}\mathcal{G}^{-1}\mathcal{G}\partial_{\sigma}\mathcal{G}^{-1}\mathcal{G}\partial_{\tau}\mathcal{G}^{-1}\right)\label{eq:flux}
\end{multline}
where $e^{0123}=+1$, $e^{1023}=-1$ etc.  The quantity (\ref{eq:flux}) is well known to those working in quantum field theory, where it appears as the divergence of the Chern--Simons current~\cite{srednicki2007}.  As also shown in~\cite{srednicki2007}, when $\mathcal{G}$ is an element of SU(2) (we shall show it is proportional to one), the above integral represents the winding number of the mapping from the closed hypersurface in $\zeta_{\mu}$ space to the three sphere representation of SU(2).  This establishes the relationship between the number of interface states in a continuous electromagnetic material and the `rotation' of the Green function at infinity (i.e. how many times the Green function explores the whole of the  SU(2) group as we explore the entire hypersurface).
\par
At first sight it appears that this integral is always zero, because the divergence of the integrand is zero
\begin{multline}
	\partial_{\mu}{\rm tr}\left[e^{\mu\nu\sigma\tau}\left(\mathcal{G}\partial_{\nu}\mathcal{G}^{-1}\mathcal{G}\partial_{\sigma}\mathcal{G}^{-1}\mathcal{G}\partial_{\tau}\mathcal{G}^{-1}\right)\right]\\
    =-{\rm tr}\left[e^{\mu\nu\sigma\tau}\left(\mathcal{G}\partial_{\nu}\mathcal{G}^{-1}\mathcal{G}\partial_{\sigma}\mathcal{G}^{-1}\mathcal{G}\partial_{\mu}\mathcal{G}^{-1}\mathcal{G}\partial_{\tau}\mathcal{G}^{-1}\right)\right]=0
\end{multline}
(the antisymmetrized product of an even number of matrices has zero trace because odd permutations are equivalent to a cyclic reordering of the indices).  By Stokes's theorem we seem to be integrating something equal to zero over the volume $V$.  However, the Green function is singular at the origin $\zeta_{\mu}=0$, which is entirely responsible for the non--zero value of (\ref{eq:flux}).  This means that the integral (\ref{eq:flux}) can be deformed into \emph{any} surface integral that encloses the origin.  A particularly useful choice is that labeled `B' in figure~\ref{fig:mode_counting}iii, where we integrate over two constant $x$ hypersurfaces
\begin{equation}
	N=\nu'(x_{\rm max})-\nu'(-x_{\rm max})\label{eq:rotated_integral}
\end{equation}
where
\begin{multline}
	\nu'(x)=\\
    -\frac{1}{24\pi^{2}}{\rm tr}\int d\zeta_0 d\zeta_2 d\zeta_3 e^{1 i j k}\left(\mathcal{G}\partial_{i}\mathcal{G}^{-1}\mathcal{G}\partial_{j}\mathcal{G}^{-1}\mathcal{G}\partial_{k}\mathcal{G}^{-1}\right)\label{eq:kxkyintegral}
\end{multline}
Performing the integrals over surfaces of constant $x$ means that---if the material is homogeneous at large $x$---we can relate the properties of wave propagation in two homogeneous media to the existence of edge states at an interface between them.  In general there is no need for the media to have any particular spatial dependence.
\par
Up to this point our treatment has been very general, only relying on the spectral symmetry of the semiclassical Green function $\mathcal{G}$, and the boundary condition that $\mathcal{G}\to0$ for large values of $\zeta_{\mu}$.  We shall return to the general case in the next section.  For the particular case of planar gyrotropic media, the field can be described by a Dirac Hamiltonian (\ref{eq:dirac_2}).  For this case, the quantity $H+{\rm i}\xi$ can be written as
\begin{align}
    H+{\rm i}\xi&=\sqrt{\boldsymbol{k}^2 k_0^{-2}+m^2+\xi^2}[{\rm i}\cos(\theta)+(\boldsymbol{n}\cdot\boldsymbol{\sigma})\sin(\theta)]\nonumber\\
    &={\rm i}\sqrt{\boldsymbol{k}^2 k_0^{-2}+m^2+\xi^2}{\rm e}^{-{\rm i}\theta\boldsymbol{n}\cdot\boldsymbol{\sigma}}
\end{align}
where we have defined
\begin{align}
	\cos(\theta)&=\frac{\xi}{\sqrt{\boldsymbol{k}^{2}k_0^{-2}+m^{2}+\xi^{2}}}\nonumber\\
	\boldsymbol{n}&=\frac{\boldsymbol{k}k_0^{-1}+m(x)\hat{\boldsymbol{z}}}{\sqrt{\boldsymbol{k}^{2}k_0^{-2}+m^{2}}}\label{eq:group_coordinates}
\end{align}
The semi--classical Green function $\mathcal{G}$ is thus clearly proportional to the $2\times2$ unitary matrix $\exp({\rm i}\theta\boldsymbol{n}\cdot\boldsymbol{\sigma})$
\begin{equation}
	\mathcal{G}=\left[H+{\rm i}\xi\right]^{-1}=\frac{-{\rm i}{\rm e}^{{\rm i}\theta\boldsymbol{n}\cdot\boldsymbol{\sigma}}}{\sqrt{\boldsymbol{k}^{2}k_0^{-2}+m^{2}+\xi^{2}}}\label{eq:green_unitary}
\end{equation}

Substituting (\ref{eq:green_unitary}) into (\ref{eq:kxkyintegral}) we find that the scalar prefactor of $1/\sqrt{\boldsymbol{k}^{2}k_0^{-2}+m^2+\xi^2}$ in (\ref{eq:green_unitary}) is irrelevant, contributing only terms with zero trace.  Using the result that
\begin{multline}
    {\rm e}^{{\rm i}\theta\boldsymbol{n}\cdot\boldsymbol{\sigma}}\partial_i{\rm e}^{-{\rm i}\theta\boldsymbol{n}\cdot\boldsymbol{\sigma}}=-{\rm i}\boldsymbol{\sigma}\cdot\bigg[\boldsymbol{n}\partial_i\theta\\
    +\sin(\theta)\left(\cos(\theta)\partial_i\boldsymbol{n}-\sin(\theta)\boldsymbol{n}\times\partial_i\boldsymbol{n}\right)\bigg]
\end{multline}
along with the general formula ${\rm tr}[(\boldsymbol{\sigma}\cdot\boldsymbol{a})(\boldsymbol{\sigma}\cdot\boldsymbol{b})(\boldsymbol{\sigma}\cdot\boldsymbol{c})]=2{\rm i}\boldsymbol{a}\cdot(\boldsymbol{b}\times\boldsymbol{c})$ we find that $\nu'$ equals
\begin{multline}
	\nu'(x)=\\
    -\frac{1}{2\pi^{2}}\int d\zeta_0 d\zeta_2 d\zeta_3 \sin^{2}(\theta)\frac{\partial\theta}{\partial \zeta_{0}}\boldsymbol{n}\cdot\left(\frac{\partial\boldsymbol{n}}{\partial \zeta_2}\times\frac{\partial\boldsymbol{n}}{\partial \zeta_3}\right)
\end{multline}
The above quantity is evidently the number of times that the unit vector $(\cos(\theta),\sin(\theta)\boldsymbol{n})$ wraps around the three sphere.  The integrand of (\ref{eq:index_integral}) is simply the Jacobian of the transformation between $\mathbb{R}^{3}$ and a point on the sphere.  Accordingly, the result of the single integral (\ref{eq:index_integral}) need not be integer (as has been found in several papers on `topological photonics' e.g.~\cite{silveirinha2015}), but the difference (\ref{eq:rotated_integral}) must be, representing the mapping from one three--sphere to another.  Performing the above integral in our case gives  
\begin{align}
	\nu'(x)&=\frac{1}{4\pi}\int d\zeta_2 \int d\zeta_3\boldsymbol{n}\cdot\left(\frac{\partial\boldsymbol{n}}{\partial \zeta_2}\times\frac{\partial\boldsymbol{n}}{\partial \zeta_3}\right)\nonumber\\
	&=\frac{1}{4\pi k_0^2}\int d^2\boldsymbol{k}\frac{m}{\left(\boldsymbol{k}^{2}/k_0^2 + m^2\right)}\nonumber\\
    &=\frac{1}{2}{\rm sign}[m(x)]
\end{align}
For the case illustrated in figure 2i of the main text,
\[
	N=\nu'(x_{\rm max})-\nu'(-x_{\rm max})={\rm sign}[m(x_{\rm max})]
\]
assuming $m(x_{\rm max})>0$ this result indicates that a single state is bound to the mirror, propagating with $k_y$ increasing as $\mathcal{E}=\sqrt{\epsilon_{1}}$ increases.  This is in agreement with the mode given by eg.~\ref{eq:edge_state} in the main text.   Thus we have proved that a winding number can be used to predict the number of interface states present between at the interface of a gyrotropic medium, at a fixed frequency $k_0$.\\[15pt]

\section{The Dirac operator for a generic medium\label{ap:chern_maxwell}}
\par
The approach developed in the main text for planar gyrotropic media can be generalized to arbitrary electromagnetic materials.  To apply formula (\ref{eq:index_integral}) we write Maxwell's equations at a fixed frequency as a first order differential equation for a $6$--vector $(\boldsymbol{E},\eta_0\boldsymbol{H})^{\rm T}$
\begin{equation}
	k_0^{-1}\left(\begin{matrix}\boldsymbol{0}&{\rm i}\boldsymbol{\nabla}\times\\-{\rm i}\boldsymbol{\nabla}\times&\boldsymbol{0}\end{matrix}\right)\left(\begin{matrix}\boldsymbol{E}\\\eta_0\boldsymbol{H}\end{matrix}\right)=\chi\cdot\left(\begin{matrix}\boldsymbol{E}\\\eta_0\boldsymbol{H}\end{matrix}\right)
\end{equation}
where the quantity $\chi$ describes all possible local linear media
\begin{equation}
\chi=\left(\begin{matrix}\boldsymbol{\epsilon}&\boldsymbol{\xi}\\\boldsymbol{\xi}^{\dagger}&\boldsymbol{\mu}\end{matrix}\right)
\end{equation}
This $6\times6$ matrix includes the permittivity $\boldsymbol{\epsilon}$ and permeability $\boldsymbol{\mu}$ tensors, and the bianisotropy $\boldsymbol{\xi}$, $\chi$ is Hermitian when the tensors $\boldsymbol{\epsilon}$ and $\boldsymbol{\mu}$ are Hermitian.  We now separate out the material tensor $\chi$ into a traceless part $\chi'$ plus a remainder 
\begin{equation}
	\chi=\frac{1}{6}{\rm tr}[\chi]1_{6}+\chi'=\mathcal{E} 1_{6}+\chi'
\end{equation}
where $\chi'$ is a fixed traceless $6\times6$ tensor.  We also introduce the angular momentum operators
\begin{align}
	\boldsymbol{L}_{1}&=\left(\begin{matrix}0&0&0\\0&0&-1\\0&1&0\end{matrix}\right),\\ \boldsymbol{L}_{2}&=\left(\begin{matrix}0&0&1\\0&0&0\\-1&0&0\end{matrix}\right),\\\boldsymbol{L}_{3}&=\left(\begin{matrix}0&-1&0\\1&0&0\\0&0&0\end{matrix}\right)
\end{align}
using these to define $6\times6$ analogues of the $\gamma$ matrices that appear in the ordinary Dirac equation (c.f.~\cite{barnett2014})
\begin{equation}
	\gamma_{i}=\left(\begin{matrix}\boldsymbol{0}&-\boldsymbol{L}_{i}\\\boldsymbol{L}_{i}&\boldsymbol{0}\end{matrix}\right)
\end{equation}
With the above definitions we can write Maxwell's equations as something reminiscent of the time independent Dirac equation
\begin{equation}
	\left[-{\rm i}k_0^{-1}\gamma_{i}\partial_{i}-\chi'\right]|\psi\rangle=\mathcal{E}|\psi\rangle\label{eq:maxwell_general_dirac}
\end{equation}
where $\langle\psi|=(\boldsymbol{E}^{\star},\eta_0\boldsymbol{H}^{\star})$.  Note however that---unlike the true Dirac equation, and that investigated in the first part of the main text---these $\gamma_{i}$ matrices do not form a Clifford algebra.
\par
For homogeneous media, the meaning of this equation is that, given a wave--vector $\boldsymbol{k}/k_0$ and the traceless part of the full material tensor $\chi$, the eigenvalue $\mathcal{E}$ is the corresponding trace of $\chi$ such that the dispersion relation is satisfied.  If we plot out the $6$ real numbers $\mathcal{E}_{n}$ as a function of $\boldsymbol{k}$ there can be gaps in each of these `spectra' (i.e. in the space of material parameters).  These gaps indicate that for this value of ${\rm tr}[\chi]/6$ fewer, or perhaps no modes are allowed to propagate in the bulk of the material, and we can again use formula (\ref{eq:index_integral}) to determine how many edge modes exist in this gap.
\par
As in the previous section, we consider a medium that is homogeneous for large values of $x$, but changes its properties over a region close to $x=0$.  It is also assumed that at infinity $\chi'$ is such that all the fields are zero.  We an thus again apply the formalism of appendix~\ref{ap:index}, determining the number of edge states from the semi--classical Green function (\ref{eq:green_wkb}), which is given by
\begin{equation}
	\mathcal{G}=\frac{1}{{\rm i}\xi+k_0^{-1}\gamma_{i}k_{i}-\chi'}=\sum_{n}\frac{|n\rangle\langle n|}{\mathcal{E}_{n}+{\rm i}\xi}\label{eq:green_eqn_chi}
\end{equation}
where the $|n\rangle$ are the solutions to (\ref{eq:maxwell_general_dirac}) for a uniform medium with the material properties $\chi'(x)$.  As is implied from equation (\ref{eq:green_wkb}), in general there is a zeroth order contribution to the spectral asymmetry $\nu$, coming purely from the semi--classical spectrum.  For simplicity we neglect this term, and the classical Hamiltonian $H=k_0^{-1}\gamma_{i}k_i-\chi'$ is assumed to have the following symmetry (an instance of which we shall see in the example below)
\begin{equation}
	THT^{\dagger}=-H\label{eq:material_symmetry}
\end{equation}
which is analogous to the chiral symmetry found e.g. in graphene~\cite{aoki2014}, and applied more widely in index theorems for the Dirac operator~\cite{weinberg1981,shiozaki2012}.  This symmetry implies that in the regions of parameter space where propagation is allowed ($\boldsymbol{k}/k_0$ real), every eigenvalue $\mathcal{E}_n$ (trace of $\chi'$) comes with a partner of the opposite sign.  Inserting expression (\ref{eq:green_eqn_chi}) into (\ref{eq:kxkyintegral}) we find
\begin{widetext}
\begin{align}
	\nu'(x)&=\frac{{\rm i}}{8\pi^{2}}\int d\zeta_0 \int d\zeta_2 \int d\zeta_3 e^{jl}\sum_{n,m}\frac{1}{(\mathcal{E}_{m}+{\rm i}\zeta_0)^{2}}\langle m|\frac{\partial H}{\partial \zeta_j}|n\rangle\frac{1}{\mathcal{E}_{n}+{\rm i}\zeta_0}\langle n|\frac{\partial H}{\partial \zeta_l} |m\rangle\nonumber\\
    &=\frac{{\rm i}}{2\pi}\int d\zeta_2 \int d\zeta_3 e^{jl}\sum_{\substack{\mathcal{E}_n> 0\\\mathcal{E}_m<0}}\frac{\langle m|\frac{\partial H}{\partial \zeta_j}|n\rangle\langle n|\frac{\partial H}{\partial \zeta_l} |m\rangle}{(\mathcal{E}_n-\mathcal{E}_m)^{2}}\nonumber\\
    &=\frac{{\rm i}}{2\pi k_0^{2}}\int dk_x \int dk_y \sum_{\substack{\mathcal{E}_n> 0\\\mathcal{E}_m<0}}\frac{\langle m|\gamma_{1}|n\rangle\langle n|\gamma_{2} |m\rangle-\langle m|\gamma_{2}|n\rangle\langle n|\gamma_{1} |m\rangle}{(\mathcal{E}_n-\mathcal{E}_m)^{2}}\nonumber\\
    &=\frac{1}{2\pi}\int dk_x \int dk_y \mathcal{F}\label{eq:general_berry_result}
\end{align}
\end{widetext}
where $e^{jl}$ is the anti--symmetric unit tensor with indices ranging over $\{2,3\}$, and we again used the coordinate system $(\zeta_0,\zeta_1,\zeta_2,\zeta_3)=(\xi,x,k_x,k_y)$.  Zero eigenvalues (i.e. zero eigenvalues at $x=\pm x_{\rm max}$) are here treated as infinitesimal positive numbers (we imagine shifting the whole spectrum up by an infinitesimal positive amount, which does not affect the value of $N$, being a difference between two values of the spectral asymmetry).  A comparison with the work of Berry~\cite{berry1984} shows that $\nu(k_y)$ is simply a sum over the Berry curvature $\mathcal{F}$ associated with the eigenfunctions of $H$ that have a negative eigenvalue, integrated over $k_x$ and $k_y$ and divided by $2\pi$.  To see this note that the sum of the Berry curvature over all the states is zero $e^{i j}\sum_{m\neq n}\langle n|\partial_i H|m\rangle\langle m|\partial_j H|n\rangle=0$ due to the ability to interchange the indices $n$ and $m$.  If the sum over $n$ runs instead only over the negative part of the spectrum then by the same argument, the sum over $m$ in the negative half of the spectrum is zero, leaving $e^{i j}\sum_{\mathcal{E}_{m}>0,\mathcal{E}_n<0}\langle n|\partial_i H|m\rangle\langle m|\partial_j H|n\rangle$, as we have in (\ref{eq:general_berry_result}).
\par
The above result is the same expression as found in the generalized Dirac operator index theorem of Fukui et al.~\cite{shiozaki2012}.  Our result shows that, through writing the Maxwell equations in the form of a Dirac operator one can prove a generalization of the known result, where the number of edge states existing in the band gap of a periodic medium can be related to the sum of the Chern numbers below the gap of interest.  However we emphasize that here the `gap' is one in the space of material parameters, not in frequency, and the theory thus has a rather different applicability.\\[15pt]
%
%
\section{Examples\label{ap:example}}

\par We now give some examples, predicting the number of interface states between inhomogeneous continuous electromagnetic media, at a fixed frequency.\\[15pt]
\noindent
%
%
\subsection{Gyrotropic media revisited\label{ap:example}}
\par
Let's reconsider the gyrotropic medium that we previously cast as a two dimensional Dirac equation.  In this case the analogous traceless material tensor $\chi'$ is given by
\begin{equation}
	\chi'=\left(\begin{matrix}-{\rm i}\alpha\boldsymbol{L}_{3}&\boldsymbol{0}\\\boldsymbol{0}&\boldsymbol{0}\end{matrix}\right)
\end{equation}
The $6\times6$ `classical' Hamiltonian is
\begin{equation}
	H=\left(\begin{matrix}{\rm i}\alpha\boldsymbol{L}_{3}&-\boldsymbol{L}_{j}k_j k_0^{-1}\\\boldsymbol{L}_{j}k_j k_0^{-1}&\boldsymbol{0}\end{matrix}\right)\label{eq:example_gyro}
\end{equation}
As assumed above, the eigenvalues of $H$ come in pairs of opposite sign, with eigenfunctions given by
\begin{widetext}
\begin{equation}
\mathcal{E}=\pm 0:\qquad|\psi\rangle=\frac{1}{\sqrt{\boldsymbol{k}^{2}+\alpha^{2}k_0^{2}}}\left(\begin{matrix}k_x \\k_y\\0\\0\\0\\-{\rm i}\alpha k_0\end{matrix}\right),\frac{1}{|\boldsymbol{k}|} \left(\begin{matrix}0\\0 \\0\\k_x\\k_y\\0\end{matrix}\right)\label{eq:gyro_eigen1}
\end{equation}
as well as,
\begin{equation}
\mathcal{E}=\pm|\boldsymbol{k}|k_0^{-1}:\qquad|\psi_{+}\rangle=\frac{1}{\sqrt{2}}\left(\begin{matrix}0 \\0\\1\\k_y/|\boldsymbol{k}|\\-k_x/|\boldsymbol{k}|\\0\end{matrix}\right),|\psi_{-}\rangle=\frac{1}{\sqrt{2}}\left(\begin{matrix}0 \\0\\1\\-k_y/|\boldsymbol{k}|\\k_x/|\boldsymbol{k}|\\0\end{matrix}\right)\label{eq:gyro_eigen2}
\end{equation}
and
\begin{equation}
\mathcal{E}=\pm\kappa k_0^{-1}:\qquad|\psi_{+}\rangle=\frac{1}{\sqrt{2}|\boldsymbol{k}|\kappa}\left(\begin{matrix}-\kappa k_y-{\rm i}\alpha k_0 k_x\\\kappa k_x -{\rm i}\alpha k_0 k_y\\0\\0\\0\\\boldsymbol{k}^{2}\end{matrix}\right),|\psi_{-}\rangle=\frac{1}{\sqrt{2}|\boldsymbol{k}|\kappa}\left(\begin{matrix}\kappa k_y-{\rm i}\alpha k_0 k_x\\-\kappa k_x -{\rm i}\alpha k_0 k_y\\0\\0\\0\\\boldsymbol{k}^{2}\end{matrix}\right)\label{eq:gyro_eigen3}
\end{equation}
\end{widetext}
where $\kappa=\sqrt{\boldsymbol{k}^{2}+\alpha^{2}k_0^{2}}$.  As demonstrated in the main text, for the second (TM) polarization there is a `gap' in the space of material parameters such that when $\mathcal{E}\in[-\alpha,\alpha]$ there is no allowed propagation.
\par
From the above expressions (\ref{eq:gyro_eigen1}--\ref{eq:gyro_eigen3}) we can calculate the Berry connection $\boldsymbol{\mathcal{A}}$ associated with the negative eigenvalues
\begin{equation}
	\boldsymbol{\mathcal{A}}={\rm i}\sum_{\mathcal{E}_{m}<0}\langle m|\boldsymbol{\nabla}_{\boldsymbol{k}}|m\rangle
\end{equation}
the curl of which is equal to (\ref{eq:general_berry_result}). To compute this quantity easily we note that the gradient of the normalization factors $N_{\boldsymbol{k}}$ in (\ref{eq:gyro_eigen1}--\ref{eq:gyro_eigen3}) provide a longitudinal contribution ${\rm i}\sum_{\mathcal{E}_m<0}\boldsymbol{\nabla}_{\boldsymbol{k}}\ln(N_{\boldsymbol{k}})$, which does not affect the value of $\mathcal{F}=\boldsymbol{\nabla}\times\boldsymbol{\mathcal{A}}$.  The zero eigenvalue solutions (\ref{eq:gyro_eigen1}) also give solely longitudinal contributions.  The negative eigenvalue solution (\ref{eq:gyro_eigen2}) is real valued, contributing zero ($\langle m|\boldsymbol{\nabla}_{k}|m\rangle=\frac{1}{2}\boldsymbol{\nabla}_{k}\langle m|m\rangle=0$).  Therefore only (\ref{eq:gyro_eigen3}) needs to be included in the value of $\boldsymbol{\mathcal{A}}$ leaving
\begin{equation}
	\boldsymbol{\mathcal{A}}=-\frac{\alpha k_0}{\kappa}\boldsymbol{\nabla}\theta_{\boldsymbol{k}}\label{eq:berry_connection_gyro}
\end{equation}
which is well defined except at the point $\boldsymbol{k}=0$.  The curl of (\ref{eq:berry_connection_gyro}) is well defined everywhere and equals
\begin{equation}
	\mathcal{F}=\frac{\alpha k_0}{\left(\boldsymbol{k}^{2}+\alpha^{2}k_0^{2}\right)^{3/2}}
\end{equation}
the total integral of which equals
\begin{equation}
	\frac{1}{2\pi}\int d^{2}\boldsymbol{k}\mathcal{F}={\rm sign}[\alpha]
\end{equation}
Using this result along with our earlier theorem relating it to the number of interface modes we find
\begin{align}
	N&=\nu'(x_{\rm max})-\nu'(-x_{\rm max})\nonumber\\
    &={\rm sign}[\alpha(x_{\rm max})]-{\rm sign}[\alpha(-x_{\rm max})]\label{eq:number_berry_integral}
\end{align}
Therefore at an interface between two gyrotropic media with opposite signs of gyrotropy $\alpha$ we can expect a total of two interface modes to cross the gap where propagation is forbidden, as the trace of the material tensor $\chi$ is increased, however the spatial dependence of $\chi$ at the interface is configured.  Meanwhile an interface between a gyrotropic media and an ordinary isotropic one (e.g. a mirror) can only lead to a single mode.  This latter case agrees with our earlier result.
\\[15pt]
\noindent
%
%
\subsection{Bianisotropic media and electric--magnetic symmetry\label{ap:example}}
\par
We can use the simple case given in the previous section to find a general class of bianisotropic materials that support one--way edge states.  Let's consider a traceless material tensor of the form
\[
	\chi'=\left(\begin{matrix}-{\rm i}\alpha\boldsymbol{L}_{3}&\boldsymbol{0}\\\boldsymbol{0}&-{\rm i}\alpha'\boldsymbol{L}_{3}\end{matrix}\right)
\]
From the previous section we know that there will be no propagation allowed in the region of parameter space $\mathcal{E}\in[-\mathcal{E}_{\rm min},\mathcal{E}_{\rm min}]$, where $\mathcal{E}_{\rm min}$ is the smallest in magnitude of $\alpha$ and $\alpha'$.  The integral of the Berry curvature equals the number of states that cross this gap and the generalization of (\ref{eq:number_berry_integral}) is straightforward
\begin{multline}
	N={\rm sign}[\alpha(x_{\rm max})]+{\rm sign}[\alpha'(x_{\rm max})]\\
    -{\rm sign}[\alpha(-x_{\rm max})]-{\rm sign}[\alpha'(-x_{\rm max})]\label{eq:edge_state_number_gyro}
\end{multline}
Now suppose we write $|\psi\rangle=U|\psi'\rangle$, where $U$ is a $6\times6$ unitary matrix.  In terms of $|\psi'\rangle$ our Dirac equation (\ref{eq:maxwell_general_dirac}) becomes
\begin{equation}
	-{\rm i}k_0^{-1} (U^{\dagger}\gamma_{i}U)\partial_{i}|\psi'\rangle-(U^{\dagger}\chi'U+{\rm i}k_0 U^{\dagger}\gamma_{i}\partial_{i}U)|\psi'\rangle=\mathcal{E} |\psi'\rangle\label{eq:transformed_dirac}
\end{equation}
The $\gamma_{i}$ matrices are invariant under the following unitary transformation (c.f.~\cite{liu2015})
\begin{equation}
	U=\left(\begin{matrix}\cos(\phi)1_{3}&\sin(\phi)1_{3}\\-\sin(\phi)1_{3}&\cos(\phi)1_{3}\end{matrix}\right)
\end{equation}
which is an instance of an electric--magnetic duality transformation, with the angle $\phi$ being position dependent.  Under this transformation (\ref{eq:transformed_dirac}) is equivalent to (\ref{eq:maxwell_general_dirac}) with the following modification of the material parameters (c.f.~\cite{liu2015})
\begin{equation}
	\chi'\to U^{\dagger}\chi'U+{\rm i}k_0^{-1} \gamma_{i}U^{\dagger}\partial_{i}U
\end{equation}
In regions where the material is homogeneous the derivative of $U$ becomes zero, and the material parameters become
\begin{widetext}
\begin{equation}
	\chi'\to\left(\begin{matrix}-{\rm i}\left(\alpha\cos^{2}(\phi)+\alpha'\sin^{2}(\phi)\right)\boldsymbol{L}_{3}&-{\rm i}(\alpha-\alpha')\sin(\phi)\cos(\phi)\boldsymbol{L}_{3}\\-{\rm i}(\alpha-\alpha')\sin(\phi)\cos(\phi)\boldsymbol{L}_{3}&-{\rm i}\left(\alpha'\cos^{2}(\phi)+\alpha\sin^{2}(\phi)\right)\boldsymbol{L}_{3}\end{matrix}\right)\label{eq:transformed_chi}
\end{equation}
\end{widetext}
Because this set of material parameters were obtained from a $\boldsymbol{k}$--independent unitary transformation, the value of the Berry connection $\boldsymbol{\mathcal{A}}={\rm i}\langle\psi|\boldsymbol{\nabla}_{\boldsymbol{k}}|\psi\rangle$ is unchanged from that of section 3a.  Therefore the formula for the number of interface states in the region of parameter space $\mathcal{E}\in[-\mathcal{E}_{\rm min},\mathcal{E}_{\rm min}]$ (\ref{eq:edge_state_number_gyro}) also holds for all the bianisotropic media given by (\ref{eq:transformed_chi}).
%
%
\section{Calculation of the reflection coefficient $r(\mathcal{E},k_y)$\label{ap:example}}
\par
In figure 3 of the main text we give a plot of the reflection coefficient and the mode profiles for a particular inhomogeneous gyrotropic medium.  This was done through reducing (\ref{eq:maxwell_general_dirac}) for the material tensor (\ref{eq:transformed_chi}) to an ordinary differential equation and using the `odeint' integration routine provided by the SciPy scientific library~\cite{scipy}.  Assuming a fixed wave--vector component $k_y$ along the $y$ axis, the optical Dirac equation (\ref{eq:maxwell_general_dirac}) reduces to a pair of equations,
\begin{multline}
    \frac{d^{2}\Psi}{d x^{2}}+\left[k_0^{2}(\mathcal{E}^{2}-\alpha^{2})-k_y^{2}+\frac{d}{dx}\left(\frac{\alpha}{\mathcal{E}}\right)k_y\right]\Psi\\
    +\frac{d}{dx}\ln\left(\frac{\mathcal{E}}{\mathcal{E}^{2}-\alpha^{2}}\right)\left(\frac{d \Psi}{dx}+k_y\frac{\alpha}{\mathcal{E}}\Psi\right)=0
\end{multline}
where $\Psi=\cos(\phi)\eta_0 H_{z}+\sin(\phi)E_{z}$, and
\begin{multline}
    \frac{d^{2}\Phi}{d x^{2}}+\left[k_0^{2}(\mathcal{E}^{2}-\alpha'^{2})-k_y^{2}+\frac{d}{dx}\left(\frac{\alpha'}{\mathcal{E}}\right)k_y\right]\Phi\\
    +\frac{d}{dx}\ln\left(\frac{\mathcal{E}}{\mathcal{E}^{2}-\alpha'^{2}}\right)\left(\frac{d \Phi}{dx}+k_y\frac{\alpha'}{\mathcal{E}}\Phi\right)=0
\end{multline}
where $\Phi=\cos(\phi)E_{z}-\sin(\phi)\eta_0 H_{z}$.  To produce the plot in figure 3a, the first of these equations was integrated with the boundary condition that on the left of the profile ($x=-X$) the wave takes the form $\phi={\rm e}^{-{\rm i}k x}$ where $k=\sqrt{k_0^{2}(\mathcal{E}^{2}-\alpha^{2})-k_y^{2}}$, and the reflection coefficient $r(\mathcal{E},k_y)$ is then given by e.g.
\begin{equation}
	r(\mathcal{E},k_y)=\frac{\Psi(X)-\frac{\rm i}{k}\Psi'(X)}{\Psi(X)+\frac{\rm i}{k}\Psi'(X)}.
\end{equation}
where the phase of $r$ depends on the arbitrary point $X$, and plays no role in our calculations.  The mode profiles were generated using the same integration procedure, numerically using the peaks in the reflection coefficient to find the parameters $\mathcal{E},k_y$ of the modes.

\newpage
\end{document}